# Highly efficient isolation of waterborne sound by an air-sealed metasurface

Xiaoxue Bai,[1] Chunyin Qiu,[1*] Xinhua Wen,[1] Shasha Peng,[1] Manzhu Ke,[1] and Zhengyou Liu[1,2,3*]

[1]Key Laboratory of Artificial Micro- and Nano-structures of Ministry of Education and School of Physics and Technology, Wuhan University, Wuhan 430072, China

[2]Institute for Advanced Studies, Wuhan University, Wuhan 430072, China

[3]Department of Physics, South China University of Technology, Guangzhou 510641, China

Underwater sound isolation has been a long-standing fundamental issue in industry and military fields. Starting from a simple theoretical model, here an air-sealed metasurface is proposed to overcome this problem. Comparing with the sample without filling air, the effective impedance of the air-sealed one is greatly reduced and strikingly mismatch with water, accompanying another merit of low sound speed. Deeply suppressed sound transmission (~$10^{-5}$) through such a metasurface is observed experimentally over a wide range of ultrasonic frequencies and incident angles.

*Author to whom correspondence should be addressed.

Email:cyqiu@whu.edu.cn; zyliu@whu.edu.cn

Traditional sound isolators are made of natural homogeneous materials, and their performance depends on the impedance contrast with the background media. It is well-known that the airborne sound can be easily isolated by a general solid slab, even for the wavelength much bigger than the slab thickness. The strong reflection stems from the extreme impedance mismatch between solid and air (a ratio usually in the order of $10^3 \sim 10^4$). Comparing with the airborne sound, a deep isolation of the underwater sound is much more difficult since a natural solid is no longer acoustically hard enough. For example, the impedance ratio between steel and water is ~30, which leads to a minimal power transmission ~0.5% even for a thick enough steel plate.

In recent decades, a variety of artificially structured materials, from bulk to flat ones, have been proposed to suppress the transmission of sound. The early representatives, phononic crystals,[1-4] are engineered with frequency gaps to forbid the sound propagation inside them. The band gap occurs in a wavelength scale comparable with the structural period, and thus an efficient sound isolation requires a sample thickness of several wavelengths. The frequency gap could be greatly lowered by the development of acoustic metamaterials,[5-8] which is produced by local resonance of the deep subwavelength building blocks. However, the applications of such bulk artificial media are often limited by the issues of bandwidth and fabrication. Recently, flat artificial structures are attracting extensive interest, because of their unusual capabilities in sound manipulations, e.g., superabsorption[9-11] and wavefront reshaping[12-16] by various acoustic metasurfaces. For airborne sound, excellent sound-proof properties[17-20] have been demonstrated by single or few layer structures designed by membranes or coiling scatterers. For waterborne sound, abnormal reflection enhancements, arising from local or collective resonances, have also been observed in grating structures.[21-25] However, the performance is limited by narrow band and unsatisfactory efficiency.

Based on a simple metasurface design, here we propose a highly efficient sound shield for water environment. The acoustic metasurface is fabricated by filling air bubbles in the apertures of a stiff grating. This reduces the impedance of the whole system into a level of air and thus forms a huge impedance ratio (~$10^3$) to water. Also, favored by the low sound speed of the air filler (comparing with the sound speed in water), the sample thickness could be tuned into a deep subwavelength. This design route has been confirmed by our ultrasonic experiments, where a broadband sound screening effect occurs for a wide range of incident angles. Besides, the superior performance of the sound isolator is insensitive to the arrangement of the air bubbles, as demonstrated by the samples of periodical and quai-periodical lattices. The lightweight acoustic metasurface is promising for underwater sound applications, e.g., noise isolations. Throughout this paper, a finite-element based software package (COMSOL MUTIPHYSICS) is adopted for numerical simulations, and the ultrasonic technique is employed to measure the transmission spectra.

In the left panel of Fig. 1(a), we present a schematic illustration for the sound shield under consideration, which consists of a rigid plate (thickness $h$) perforated with a square array (lattice constant $a$) of circular holes (radii $r$). It has been proved that,[26-28] when the wavelength is much larger than the structural period, such a grating behaves effectively as a strongly anisotropic fluid slab of thickness $h$ [right panel in Fig. 1(a)], which can be characterized by a scalar bulk modulus $\kappa_e = \kappa_f / f$ and a mass density tensor $\boldsymbol{\rho}_e = \mathrm{diag}[\rho_{ex}, \rho_{ey}, \rho_{ez}]$ with $\rho_{ex} = \rho_{ey} \to \infty$ and $\rho_{ez} = \rho_f / f$. Here $f$ is the area filling ratio of the holes $f = \pi r^2 / a^2$, and $\kappa_f$ and $\rho_f$ are respectively the bulk modulus and mass density of the fluid filled in the holes. For such an anisotropic slab immersed in a background fluid (i.e., water here), the transmission coefficient at normal incidence can be simply evaluated by the formula[28]

$$t = \frac{4uZ_r}{(Z_r+1)^2 - u^2(Z_r-1)^2}, \qquad (1)$$

where $u = e^{ik_f h}$ is a phase factor, $k_f$ is the wavenumber of the hole filler, and $Z_r = Z_w/Z_e$ is the impedance ratio between water ($Z_w = \rho_w c_w$) and the effective medium ($Z_e = \rho_f c_f / f$). Here $\rho$ and $c$ represent mass density and sound speed, and the subscripts "$w$" and "$f$" label the parameters of water and the hole filler.

Therefore, only a moderate impedance ratio ($Z_r = f$) is obtained if the hole is filled by water, whereas a remarkable impedance mismatch emerges if air is filled instead, because of its extreme density contrast to water. This enables a good solution for the waterborne sound isolation: a minimal power transmission $|t|^2 \approx 4/Z_r^2$, occurring at the wavelength with $u^2 = -1$. Besides, the effective sound speed penetrating the hybrid structure depends on the air filler in the holes, of which the sound speed is much slower than water. This gives another important merit, i.e., shrinking the sample thickness of the device.

Note that the physics mechanism behind the sound-proof is much different from the system consisting of freely distributed bubble clusters in water,[29,30] where the local resonances of air bubbles contribute to the excellent screening effect in the low frequency. Recently, a similar idea is also used to realize the so-called superabsorption[31] by filling bubbles in a layer of soft solid medium. In those cases, the media immediately enclosing the bubbles are easily compressed and thus enable low frequency Mie resonances. In this work, the bubbles fixed by the stiff frame are only squeezable in the thickness direction, which allow only Fabry-Perot (FP) resonances at much higher frequencies than the aforementioned Mie resonances. A collection of the fixed bubbles will effectively form a flat wall of extremely low impedance, and strongly reflect the underwater sound except near the FP resonant frequency.

To confirm the above idea, in Fig. 1(b) we present the transmission spectra for acoustic waves incident normally onto a rigid grating with structural parameters $h = 3.1mm$, $r = 0.6mm$ and $a = 2mm$ (thus the filling ratio $f = 0.283$). The holes are filled with water or air. The acoustic parameters used are: densities $10^3 kg/m^3$ (water) and $1.29 kg/m^3$ (air), and sound speeds $1.49 km/s$ (water) and $0.34 km/s$ (air). In this structure geometry, the impedance ratio $Z_r$ can be estimated as high as ~960 for the air-filled system. As observed in Fig. 1(b), as a whole, both the theoretical prediction and the full-wave simulation demonstrate a much lower transmission for the air-filled system, except for the frequency in the vicinity of the sharp peak. In addition to the periodical peaks originated from the FP resonance (with $u^2 = 1$), asymmetric Fano-like peaks arising from certain intrinsic hole modes also emerge in the high frequency region, as exemplified in Fig. 1(c). These resonances are less frequently observed for the system without filling air, since the wavelength in water is much longer than that in air. It is worth pointing out that, here the excellent sound isolation persists even for the wavelength (in water) comparable with the lattice constant, although our model starts from the effective medium theory.

Now we experimentally check the performance of the sound isolation for the above metasurface of square lattice. Figure 2(a) shows the photos of the real samples filled without (left panel) and with (right panel) air bubbles. To stably fix the air bubbles in the holes when the sample immersed in water, two (acoustically transparent) thin films (of thickness ~10um) are closely glued on the both sides of the sample. Comparing with the underwater sound shield formed by free-standing gas bubbles,[29,30] the treatment here avoids the complex bubble dynamics easily. Figure 2(b) illustrates our experimental setup. In our experiment, the sample is closely clamped and placed in between a pair of identical transducers, one for generating sound and the other for receiving the signal. The sample can be inclined to a desired angle for realizing oblique incidence. The diameters of the transducers are $25mm$. The distances from the transducers (aligned collinearly with each other) to the sample are both $15cm$, which are far enough to obtain an approximate planar wavefront. The entire assembly is immersed in a big water tank. The power transmission is obtained through the well-known ultrasonic transmission technique.

In Fig. 3(a) we present the power transmission measured at normal incidence. As anticipated, the sound transmission through the sample immersed directly in water is considerably high as a whole, associated with a minimal power transmission ~0.1. Once air is sealed in the holes, the transmission is strikingly suppressed in the full spectrum. The sharp peaks predicted in Fig. 1(b) are not captured by the frequency resolution of the spectra. The energy dissipation of sound (not involved in our model) is probably sizable near the FP or Fano resonances (where the sound is penetrable and strongly localized in the air bubbles), in contrast to the negligible absorption away from the resonances because of the nearly perfect reflection. It is worth pointing out that, the broadband deep sound isolation will be kept for a wide range of incident angles, considering the fact that the impedance ratio changes simply by a factor of $1/\cos\theta$ in Eq. (1),[28] where $\theta$ is the incident angle measured from the normal of the sample. This is confirmed in Fig. 3(b) by $\theta=10^\circ$, $30^\circ$ and $45^\circ$.

As mentioned in our theoretical model, the excellent sound isolation stems from the great mismatch between the impedances of the effective medium and water. Therefore, the phenomenon depends weakly on the structure lattice, and even remains for a sample with randomly distributed holes. This property will be beneficial in fabricating real samples. To exemplify this character, we consider a sample made of a steel plate (of thickness $h=0.6mm$) perforated with a two-dimensional eightfold quasiperiodic array of subwavelength holes ($r=0.25mm$). As depicted in Fig. 4(a), the lattice is generated by a dual grid method, where the two basic units (i.e. the square and the rhombus of vertex angle 45°) are featured by the side length $d=3mm$. Figure 4(b) shows the transmission spectra for the samples incident at different angles. Comparing with the result for the water-filled sample, the power transmission through the air-sealed one is deeply suppressed again, despite of the deep subwavelength thickness of the metasurface sample (~1/5 wavelength in water even at the highest frequency under consideration). Note that the transmission peak centered at 0.287 MHz corresponds to the first FP resonance in the air disk, which is incidence insensitive due to the effectively strong anisotropy of the air-filled sample. Different from the case in Fig. 3 (with sample thickness $h=3.1mm$), here the FP peak is clearly observed since it is much broader for this thinner sample ($h=0.6mm$).

In conclusion, we have proposed a metasurface approach to achieve highly efficient sound isolation in water environment. The sample is simply fabricated by a stiff grating sealed with gas in the apertures. A broadband deep suppression of the sound transmission has been experimentally observed for a wide range of incident angles, even for the sample with a thickness of deep subwavelength. Many applications can be anticipated for such lightweight sound isolators. For example, one may fabricate a cage-like sound shield to cut off sound communication with the outside world, which could be useful in underwater military.

**Acknowledgements**

This work is supported by the National Basic Research Program of China (Grant No. 2015CB755500); National Natural Science Foundation of China (Grant Nos. 11374233, 11534013, and 11504275).

**Figures and Figure Captions:**

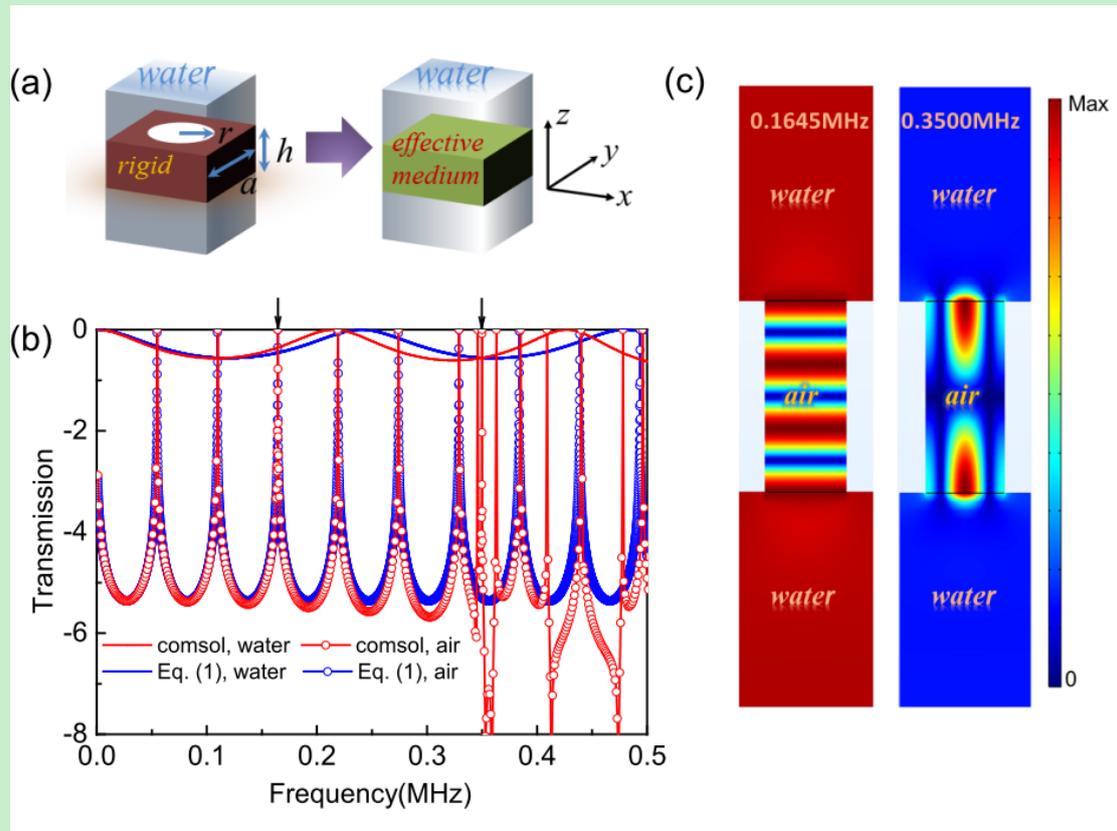

FIG. 1. (a) Left panel: a rigid plate drilled with a square lattice array of holes, where $a$, $r$ and $h$ describe the lattice constant, radii of the holes and thickness of the plate, respectively. The background fluid is water and the holes can be filled with water or air. Right panel: an effective anisotropic fluid slab immersed in water. (b) Simulated transmission spectra (in $\log_{10}$ scale) for acoustic waves incident normally onto the water-filled or air-filled metascreens, comparing with the analytical data for the corresponding effective fluid slabs. (c) Simulated pressure amplitude distributions at the resonant frequencies 0.1645 MHz and 0.3500 MHz [depicted by arrows in (b)]. The left panel shows an amplitude pattern of typical FP resonance in the air-filled hole, whereas the right panel reveals a resonance of the high order wave guiding mode in the hole (see the nonuniform pressure amplitude in the cross section of the hole, which cannot be captured by the effective medium model).

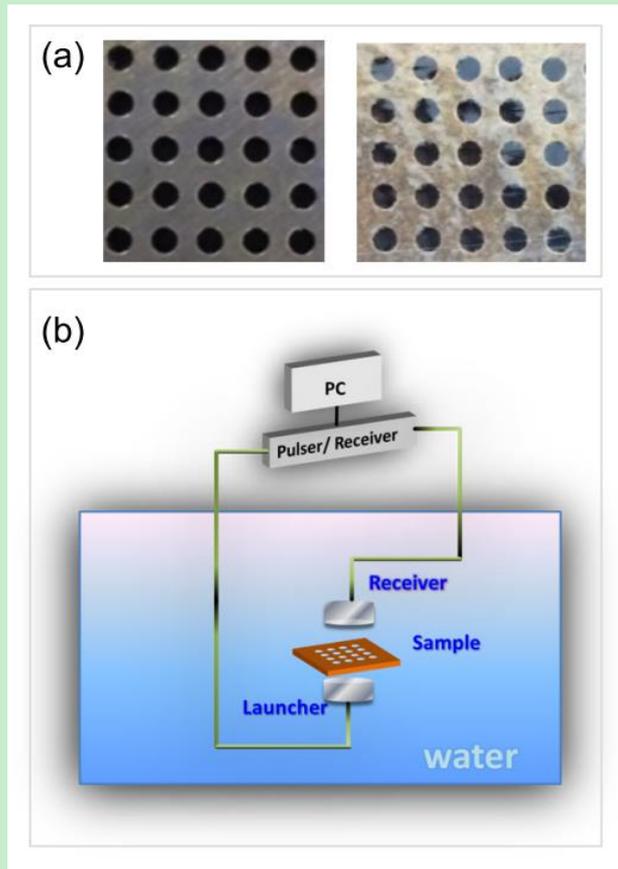

FIG. 2. (a) Left panel: Photo of a brass plate (of thickness $h=3.1mm$) drilled with a square array of circular holes. The structural parameters are identical to those involved in Fig. 1. Right panel: the metasurface with air sealed in the holes. (b) A schematic view of the experimental setup.

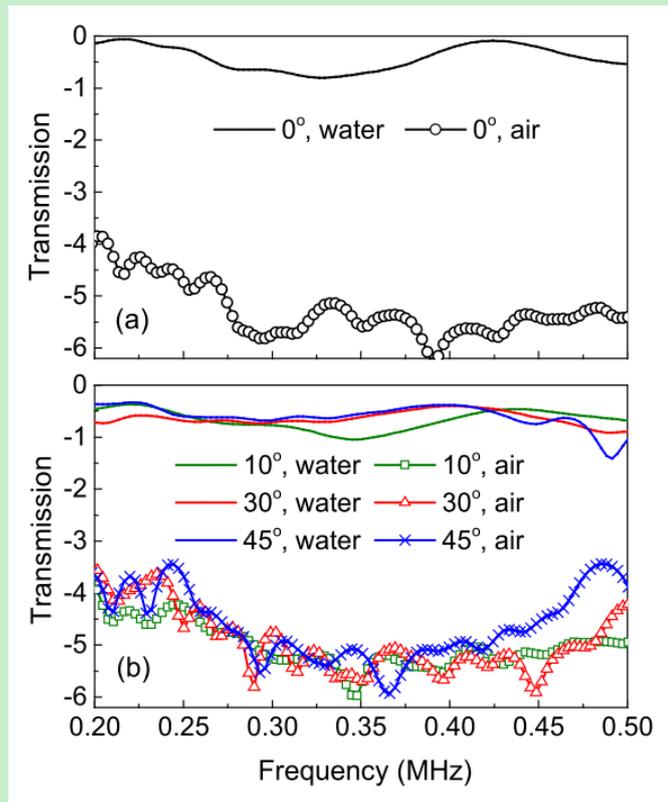

FIG. 3. (a) Experimental power transmission at normal incidence for the sample sealed with air bubbles in the circular holes, comparing with the data for the sample directly immersed in water. (b) The same as (a), but for three oblique incident angles.

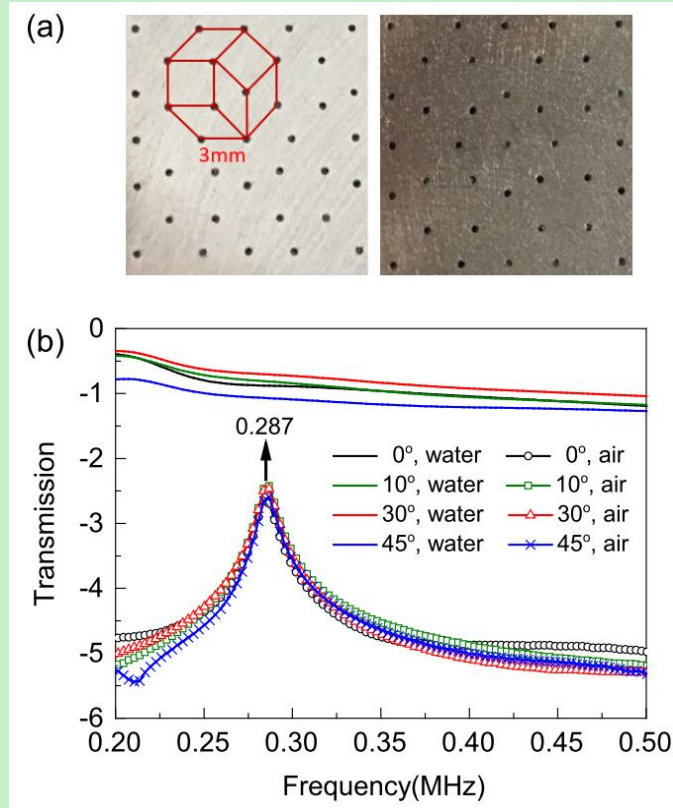

FIG. 4. (a) Photo of a steel plate (of thickness $h=0.6mm$) perforated with an eightfold quasiperiodic array of circular holes at the vertices of the square and rhombus tiles, without (left panel) and with (right panel) filling air bubbles in the holes. (b) Experimental transmission spectra for the acoustic waves incident onto the samples with and without air bubbles. The data measured at all incident angles indicate a deeply suppressed sound transmission for the air-filled sample, except near the frequency of FP resonance.